\documentclass[conference]{IEEEtran}
\usepackage{color,soul}
\usepackage{epstopdf}
\IEEEoverridecommandlockouts
\usepackage{cite}
\usepackage{amsmath,amssymb,amsfonts}
\usepackage{algorithmic}
\usepackage{graphicx}
\usepackage{textcomp}
\usepackage{xcolor}
\def\BibTeX{{\rm B\kern-.05em{\sc i\kern-.025em b}\kern-.08em
    T\kern-.1667em\lower.7ex\hbox{E}\kern-.125emX}}
\begin{document}
\title{Around-the-corner Radar Sensing Using Reconfigurable Intelligent Surface\\}
\author{\IEEEauthorblockN{Kainat Yasmeen, Debidas Kundu, and Shobha Sundar Ram}
\IEEEauthorblockA{\textit{Department of Electronics and Communication Engineering} \\
\textit{Indraprastha Institute of Information Technology}\\
\textit{} New Delhi, India \\
kainaty@iiitd.ac.in, debidas@iiitd.ac.in, shobha@iiitd.ac.in}}
\maketitle
\begin{abstract}
Around-the-corner radar (ACR) sensing of targets in non-line-of-sight (NLOS) conditions has been explored for security and surveillance applications and look-ahead warning systems in automotive scenarios. Here, the targets are detected around corners without direct line-of-sight (LOS) propagation by exploiting multipath bounces from the walls. However, the overall detection metrics are weak due to the low strength of the multipath signals. Our study presents the application of reconfigurable intelligent surface (RIS) to improve radar sensing in ACR scenarios by directing incident beams on the RIS into NLOS regions. Experimental results at 5.5 GHz demonstrate that micro-Doppler signatures of the walking motion of humans can now be captured in NLOS conditions through the strategic deployment of RIS. 
\end{abstract}


\section{Introduction}
\label{sec:Introduction}
Over several decades, radar systems have been developed and studied for detecting, tracking, and localizing targets in non-line-of-sight (NLOS) scenarios. This included ground penetration radars for detecting buried objects \cite{cassidy2009ground}, through-wall radar for security and surveillance applications \cite{ram2008doppler}, and around-the-corner radars (ACR), which have been recently explored for surveillance applications in urban environments and for look-ahead warning systems in automotive applications\cite{soldovieri2016exploitation,rabaste2019detection}. Urban environments are characterized by walls and buildings blocking the targets from the radar's view. At low microwave frequencies (below X-band), electromagnetic signals penetrate through the wall materials supporting through-wall radar sensing. However, due to the low skin depth of building materials at higher frequencies, the signals undergo multipath scattering along the surface of the walls rather than through-wall propagation \cite{ram2010simulation,vishwakarma2020micro}. Hence, at these frequencies, around-the-corner sensing based on the exploitation of multipath signals is being researched for detecting targets \cite{li2021multiple}. However, there are some limitations associated with this approach. First, the multipath signals are often weak, resulting in poor detection metrics. Second, when the signal undergoes multiple bounces, the localization algorithms yield several possible outcomes since the uniqueness criteria are no longer satisfied.
More recently, both limitations have been addressed using a reconfigurable intelligent surface (RIS) for NLOS radar sensing.

An RIS consists of a two-dimensional array of reflective elements and operates with low energy consumption, making it a largely passive device. In an RIS, the phase, amplitude, and frequency of the reflected or refracted waves can be controlled by reconfiguring the incident transmission with the passive elements \cite{basar2019wireless}. This capability provides additional flexibility and modification for wireless channel impulse response and effective system optimization. An RIS can be installed outside on building exteriors or interior ceilings or walls. By dynamically tuning the characteristics of the reflected signals, the RIS can enhance communication performance, increase spatial coverage, and improve signal quality in challenging environments \cite{liu2021reconfigurable}. 
Recently, RISs have demonstrated effectiveness in various other applications, such as wireless energy transfer \cite{zhao2020wireless}, positioning and mapping \cite{wymeersch2020radio}, and integrated sensing and communication.\\
\indent Very recently, RIS-aided radar sensing has also been investigated for a practical, low-cost, and energy-efficient approach to enhance the sensing capabilities \cite{liu2021reconfigurable,buzzi2021radar,buzzi2022foundations,mercuri2023reconfigurable}. In \cite{buzzi2021radar}, 
In \cite{buzzi2021radar}, a theoretical analysis was conducted on radar and RIS configurations with varying spacing, demonstrating that RIS can substantially enhance the effectiveness of radar target detection.
\cite{buzzi2022foundations} investigates the performance boost a RIS provides to a multiple input multiple output radar. In \cite{mercuri2023reconfigurable}, an RIS was utilized to improve radar-based indoor localization by mitigating multipath effects. In an ACR scenario, the RIS enables the propagation of signals to reach NLOS regions without relying on multipath signals. The main advantage of using RIS for this purpose is that it enhances the overall detection metrics due to lower path loss than that encountered in traditional ACR based on multipath. Additionally, due to the focused transmission of beams by the RIS, the uniqueness criteria are satisfied, potentially enhancing the localization estimation accuracy. Existing works have focused on theoretical studies of RIS in ACR scenarios. However, practical considerations such as the limited bandwidth and beamwidth of the RIS were not factored into the overall system performance. This work aims to overcome these limitations by studying RIS-aided ACR sensing using measurement data collection in real-world conditions. \\  
\indent For the purposes of this study, an RIS of $256\times160$ mm dimensions with $16\times10$ unit cells has been developed. It uses a broadband 1-bit coding metasurface unit cell loaded with PIN diode. The unit cell provides a fractional bandwidth (FBW) of 18.51$\%$ with the center frequency at 5.45 GHz for the $180^0$ $\pm 20^0$ phase difference between the ON and OFF states. The designed RIS is operated at 5.5 GHz in this study. This RIS is deployed with a narrowband monostatic radar in an ACR scenario within a research building premises to detect and track human motions. The radar data are processed to obtain joint time-frequency spectrograms in free space, ACR, and RIS-aided ACR scenarios. The results show both the promise and challenges of using RIS with radar.\\
\indent The organization of this paper is as follows: the research methodology is presented in Section.\ref{sec:researchmethodology}. The results are discussed in Section.\ref{sec:Result} while the future scope of the work is discussed in the conclusion in Section.\ref{sec:conclusion}.

\section{Research Methodology}
\label{sec:researchmethodology}
This work aims to enhance radar target detection in NLOS conditions with the assistance of RIS. We have considered a scenario where the radar transmits and receives through a beam directed at the RIS. The objective is for the RIS to focus the incoming wavefront from the transmitter onto the target. It then reflects the scattered returns from the target back to the receiver.

\subsection{Problem Statement}
\begin{figure}[htbp]
\centering
\includegraphics[scale=0.39]{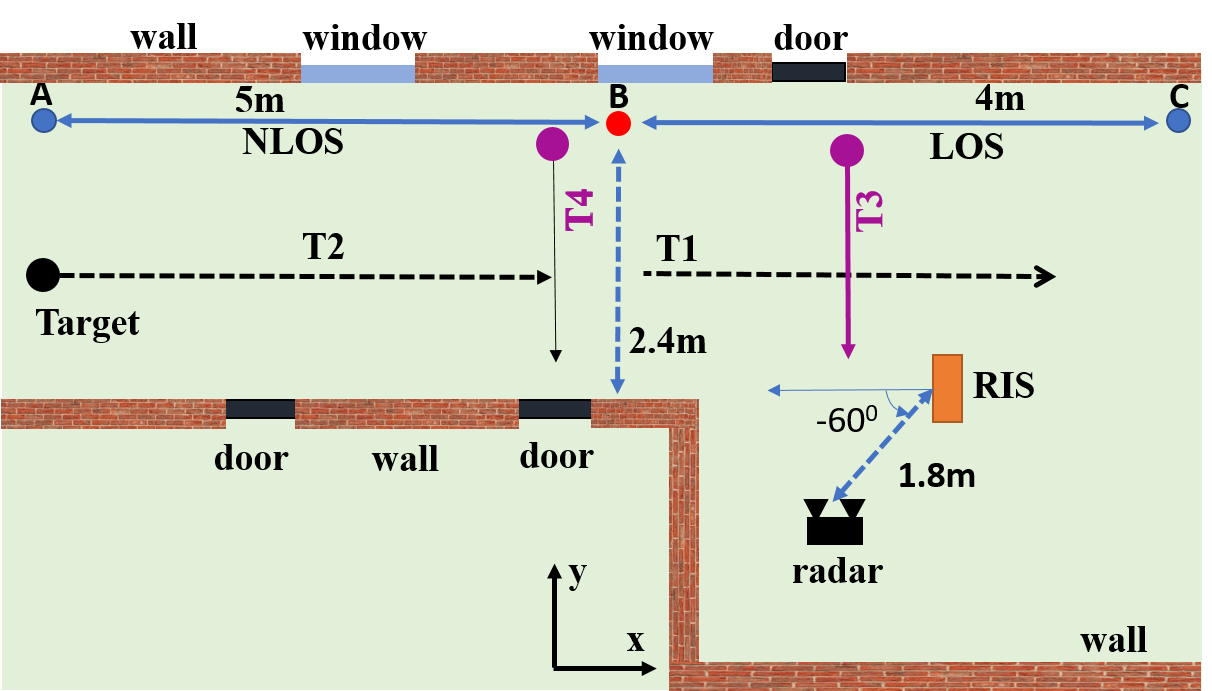}
\caption{Illustration of radar setup for around-the-corner target detection using RIS.}
\label{fig:illustration}
\end{figure}

We have considered an L-shaped corridor, which is 2.4 m wide with 30 cm brick walls, doors, and windows, as shown in Fig.\ref{fig:illustration}.
The monostatic radar antennas are oriented toward the positive y-axis in this configuration.
Due to the presence of the wall, the region between A and B is NLOS with respect to the radar, while the area between B and C is in the LOS of the radar, based on the antenna beamwidth. An RIS is strategically placed at 1.8 m from the radar with an incidence angle of $-60^0$ with respect to the negative x-axis such that the reflected wave is directed towards the NLOS direction. Both the radar and RIS are positioned at a height of 1.1 m. 
 
In this scenario, the target is a human walking with an average speed of 1.5 m/s. 
We examined four walking trajectories: (i) T1: This motion is tangential to the radar within the radar LOS, moving from point B towards C; (ii) T2: This is a walking motion along the positive x-axis from A to B within the radar NLOS; (iii) T3: This is a radial motion in front of the radar in the LOS region; and (iv) T4: Here the target moves along the negative y-axis within the NLOS region of the radar. 
 
\subsection{Experimental Setup}
\begin{figure*}[!t]
\vspace{0.2cm}
\centering
\includegraphics[width=155mm, height=73mm]{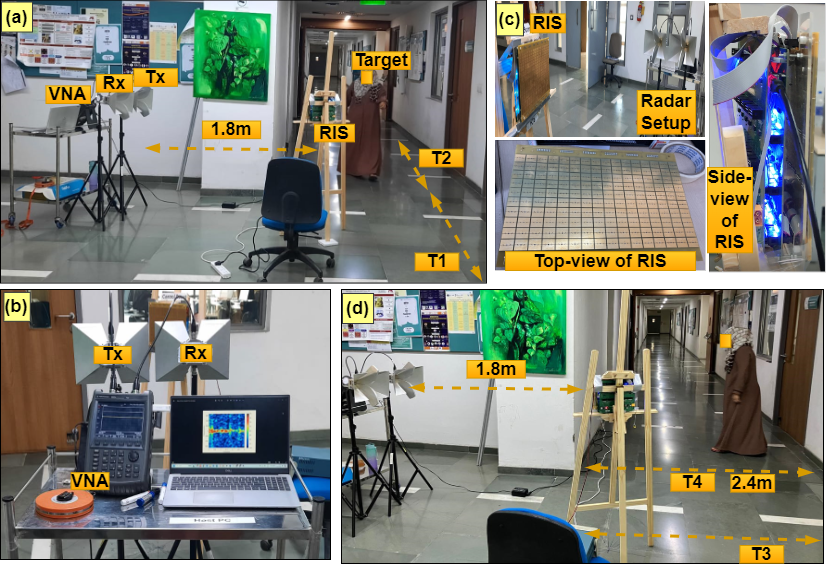}
\caption{Experimental radar setup for around-the-corner target detection using RIS : (a) trajectory T1 and T2, (b) close-up view of radar setup, (c) close-up view of RIS, and (d) trajectory T3 and T4.}
\label{fig:Experimental_setup}
\end{figure*}

We configured a FieldFox N9926A vector network analyzer (VNA) as a monostatic radar to collect measurement data as shown in Fig.\ref{fig:Experimental_setup}a. 
The VNA performs two-port scattering parameter ($S_{21}$) measurements in the time domain, centered at 5.5 GHz with a bandwidth of 10 kHz. Two HF907 horn antennas are connected to its ports. The VNA transmits at +3 dBm power, and the sampling frequency is set to the maximum limit of 370 Hz to avoid aliasing issues during Doppler measurements.
Each measurement data collection lasts 3 seconds, resulting in approximately 4000 samples. Figure \ref{fig:Experimental_setup}b provides a close-up view of the radar setup.

The unit cell of the RIS is designed using the technique given in \cite{10463937}. It has a periodicity of 16 mm. The board comprises an array of $16\times10$ unit cells and is operated at 5.45 GHz. From theoretically calculated and full-wave simulated results, it is observed that the 3-dB beamwidth of the RIS varies from $11^0$ to  $12^0$ when the reflected beam scans from $\theta_r = 15^\circ$ to $\theta_r = 60^\circ$ for the oblique incident beam with $\theta_i = -60^\circ$. A close-up view of the RIS board used for data collection is shown in Fig.\ref{fig:Experimental_setup}c.

In the ACR scenario, data collection is performed using the experimental setup represented in Fig.\ref{fig:Experimental_setup}
d. Firstly, the data is collected in the absence of RIS. 
Next, we introduce the RIS and repeat the measurements. In Fig.\ref{fig:Experimental_setup}a, the target moves tangentially to the radar and towards the RIS. In Fig.\ref{fig:Experimental_setup}d, the target moves tangentially to the RIS at a distance of 2 m, walking back and forth in the width of the corridor.
 In each case, the VNA captures the scattered signal from the target, amplifies it, demodulates it in phase-quadrature, and digitizes it. The resulting complex measurement data are then gathered and transferred to a laptop for further analysis. Next, we apply the short-time Fourier transform (STFT) to the time-domain radar data to generate micro-Doppler spectrograms, as shown in
\begin{equation}
  \textbf{STFT}(t,f) = \int_{} x(\tau)h(t-\tau)e^{-j2\pi f\tau}d\tau, 
\end{equation}
where \( h(t) \) is the window function with a duration of 0.3 seconds used for short-time analysis\cite{ram2008simulation}. 

\section{Result and Discussion}
\label{sec:Result}
\begin{figure}[htbp]
\centering
\includegraphics[scale=0.35]{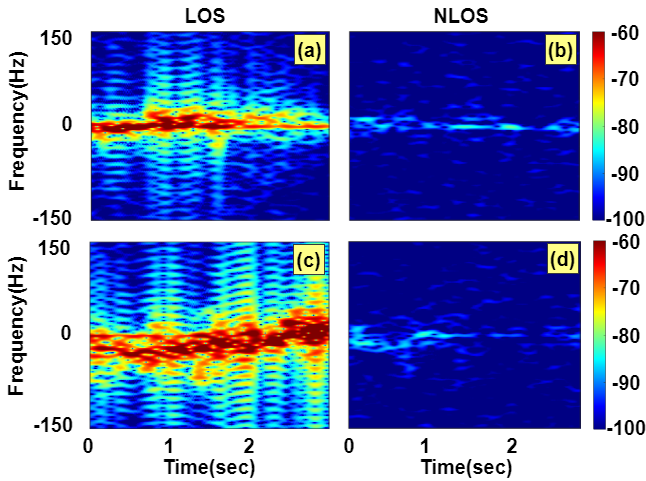}
\vspace{-1mm}
\caption{Spectrograms of a human walking under different conditions captured using around-the-corner radar at 5.5 GHz: (a) and (c) shows trajectories T1 and T3 in LOS conditions, respectively, while (b) and (d) show trajectories T2 and T4 in NLOS conditions.}
\label{fig:LOS_NLOS}
\end{figure}

In this section, experimental results are provided with detailed analysis and discussion. The micro-Doppler signatures of the human motions in the absence of RIS are depicted in Fig.\ref{fig:LOS_NLOS}. 
As illustrated in Fig.\ref{fig:LOS_NLOS}a, the human walks tangentially to radar along trajectory T1 in LOS. Here, we observe the torso generates strong returns, but the micro-Doppler from the swinging arms and legs is weaker. In a tangential motion, the overall Doppler offsets are low due to the nature of the motion. The target scattered returns from the human following T2 trajectory under NLOS conditions are shown in Fig.\ref{fig:LOS_NLOS}b. The spectrogram signals are weaker due to radar signals returning via multipath instead of a direct path, as in LOS conditions. The micro-Doppler signatures of a human walking towards the radar along trajectories T3 and T4 are depicted in Fig.\ref{fig:LOS_NLOS}c and Fig.\ref{fig:LOS_NLOS}d.
In Fig.\ref{fig:LOS_NLOS}c, the strength of the returns is stronger again due to the propagation of signals in LOS conditions. Also, the overall Doppler offsets are more significant than the tangential motion. When a human moves toward the radar, the Doppler shifts are positive. When the human turns, it becomes zero, and when the human moves away, it becomes negative. In contrast, Fig.\ref{fig:LOS_NLOS}d corresponds to NLOS conditions, where the strength of radar signatures is weak. The absence of detectable patterns in the NLOS micro-Doppler signatures makes observing human movement around obstacles challenging.
\begin{figure*}[htbp]
\centering
\includegraphics[width=150mm, height=67mm]{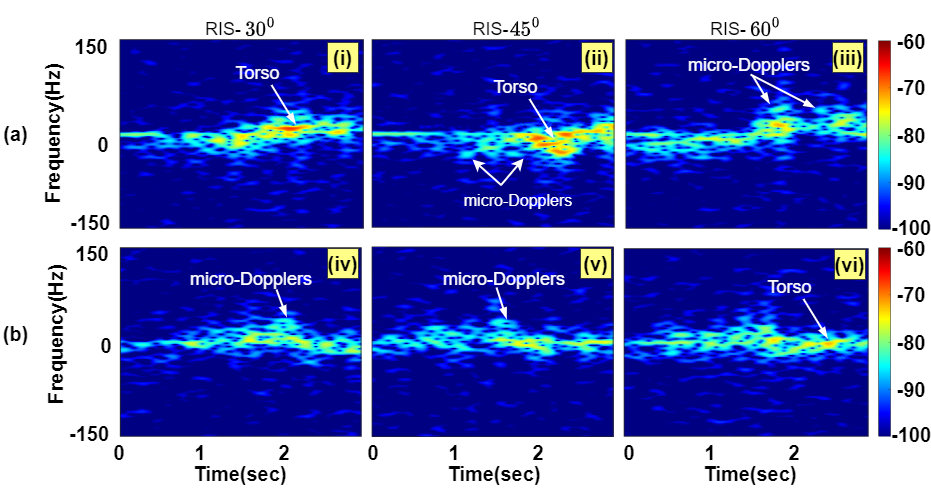}
\caption{Spectrograms of walking motion in NLOS conditions using RIS: (a) trajectory T2 and (b) trajectory T4 for RIS angles of $30^0$, $45^0$, and $60^0$, respectively.}
\label{fig:spect_with_RIS}
\end{figure*}
Next, we repeat the measurements after introducing an RIS as described before.
In Fig.\ref{fig:spect_with_RIS}, the micro-Doppler signatures for NLOS conditions when RIS is kept at an incident angle of $-60^0$ with respect to radar are presented. The RIS can switch the beam to different reflection angles. Hence, we have considered three different observation/reflection angles, i.e., $30^0$, $45^0$, and $60^0$, respectively, for this experiment. In the first row of Fig.\ref{fig:spect_with_RIS}a, depicted the trajectory T2 where human is walking along the tangential path to the radar towards the RIS from a distance of 5 m. We observe that the micro-Doppler signatures in NLOS conditions are observed with the incorporation of RIS. The strength of the returns is weaker compared to free space, but it is more robust than the returns in the NLOS case without RIS.
At 5.5 GHz frequency, we observe that Doppler returns are detected between 1 s and 3 s for RIS-$30^0$ and RIS-$45^0$ when the human is positioned closer to the RIS. 
However, In the second row in Fig.\ref{fig:spect_with_RIS}b, it is observed that the micro-Doppler signatures in NLOS are noticeable. 
This demonstrates that the RIS-assisted propagation mechanism is effective at this frequency. 
It is important to note that the signal undergoes a two-way transmission path through the RIS. This results in greater attenuation compared to one-way propagation in communication scenarios. 
Additionally, some Doppler spreading is detected due to the radar signals reflecting off the walls and creating multipath returns.
\begin{figure}[htbp]
\centering
\includegraphics[scale=0.31]{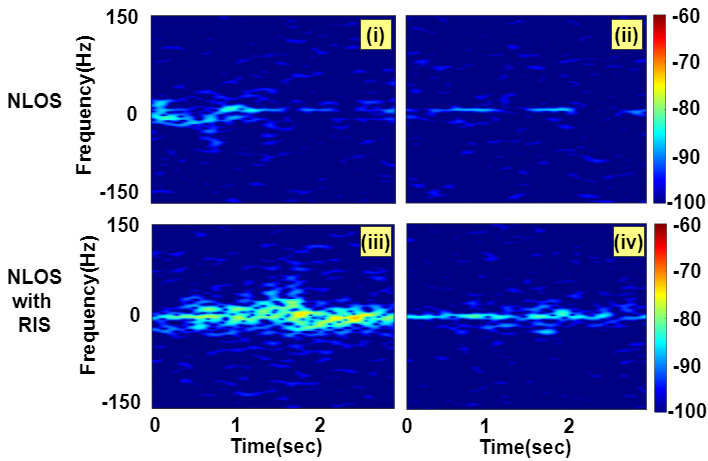}
\vspace{-1mm}
\caption{Spectrograms of a walking motion on trajectory T4 in NLOS conditions: (i) and (ii) without RIS, (iii) and (iv) with RIS.}
\label{fig:path_2_2m_3m}
\end{figure}
Figure \ref{fig:path_2_2m_3m} depicts the results of the human walking on trajectory T4, positioned 2 meters and 3 meters away from the RIS in columns 1 and 2, respectively. We observe that when the target is closer to the RIS, human motion is detected more effectively. However, as the distance increases, the reflected power decreases. Therefore, expanding the RIS field of view is recommended for future improvements to enhance detection at longer ranges.
\section{Conclusion}
\label{sec:conclusion}
This paper explores the application of RIS for target detection in around-the-corner radar (ACR) scenarios. By leveraging RIS's scattering properties, we explore how it can enhance radar detection capabilities, thus advancing NLOS radar detection. The experimental analysis shows that the performance is limited by the RIS's field of view. Also, these results were generated with narrowband radar data. 
Therefore the future scope of the work is to expand the RIS's field of view and to study its performance on wideband radar signatures in single and multiple target scenarios. 
\bibliography{Bibliography}

@article{basar2019wireless,
  title={Wireless communications through reconfigurable intelligent surfaces},
  author={Basar, Ertugrul and Di Renzo, Marco and De Rosny, Julien and Debbah, Merouane and Alouini, Mohamed-Slim and Zhang, Rui},
  journal={IEEE access},
  volume={7},
  pages={116753--116773},
  year={2019},
  publisher={IEEE}
}

@article{rabaste2019detection,
  title={Detection--localization algorithms in the around-the-corner radar problem},
  author={Rabaste, Olivier and Bosse, Jonathan and Poullin, Dominique and S{\'a}enz, Israel D Hinostroza and Letertre, Thierry and Chonavel, Thierry and others},
  journal={IEEE Transactions on Aerospace and Electronic Systems},
  volume={55},
  number={6},
  pages={2658--2673},
  year={2019},
  publisher={IEEE}
}

@article{soldovieri2016exploitation,
  title={Exploitation of ubiquitous Wi-Fi devices as building blocks for improvised motion detection systems},
  author={Soldovieri, Francesco and Gennarelli, Gianluca},
  journal={Sensors},
  volume={16},
  number={3},
  pages={307},
  year={2016},
  publisher={MDPI}
}

@article{ram2008doppler,
  title={Doppler-based detection and tracking of humans in indoor environments},
  author={Ram, Shobha Sundar and Li, Yang and Lin, Adrian and Ling, Hao},
  journal={Journal of the Franklin Institute},
  volume={345},
  number={6},
  pages={679--699},
  year={2008},
  publisher={Pergamon}
}

@inproceedings{ram2008simulation,
  title={Simulation of human microDopplers using computer animation data},
  author={Ram, Shobha Sundar and Ling, Hao},
  booktitle={2008 IEEE Radar Conference},
  pages={1--6},
  year={2008},
  organization={IEEE}
}

@inproceedings{vishwakarma2020micro,
  title={Micro-Doppler signatures of dynamic humans from around the corner radar},
  author={Vishwakarma, Shelly and Rafiq, Aaquib and Ram, Shobha Sundar},
  booktitle={2020 IEEE International Radar Conference (RADAR)},
  pages={169--174},
  year={2020},
  organization={IEEE}
}

@article{buzzi2021radar,
  title={Radar target detection aided by reconfigurable intelligent surfaces},
  author={Buzzi, Stefano and Grossi, Emanuele and Lops, Marco and Venturino, Luca},
  journal={IEEE Signal Processing Letters},
  volume={28},
  pages={1315--1319},
  year={2021},
  publisher={IEEE}
}

@article{buzzi2022foundations,
  title={Foundations of MIMO radar detection aided by reconfigurable intelligent surfaces},
  author={Buzzi, Stefano and Grossi, Emanuele and Lops, Marco and Venturino, Luca},
  journal={IEEE Transactions on Signal Processing},
  volume={70},
  pages={1749--1763},
  year={2022},
  publisher={IEEE}
}

@article{mercuri2023reconfigurable,
  title={Reconfigurable intelligent surface-aided indoor radar monitoring: A feasibility study},
  author={Mercuri, Marco and Arnieri, Emilio and De Marco, Raffaele and Veltri, Pierangelo and Crupi, Felice and Boccia, Luigi},
  journal={IEEE Journal of Electromagnetics, RF and Microwaves in Medicine and Biology},
  year={2023},
  publisher={IEEE}
}

@article{liu2021reconfigurable,
  title={Reconfigurable intelligent surfaces: Principles and opportunities},
  author={Liu, Yuanwei and Liu, Xiao and Mu, Xidong and Hou, Tianwei and Xu, Jiaqi and Di Renzo, Marco and Al-Dhahir, Naofal},
  journal={IEEE communications surveys \& tutorials},
  volume={23},
  number={3},
  pages={1546--1577},
  year={2021},
  publisher={IEEE}
}

@article{zhao2020wireless,
  title={Wireless power transfer empowered by reconfigurable intelligent surfaces},
  author={Zhao, Long and Wang, Zhouyin and Wang, Xiaodong},
  journal={IEEE Systems Journal},
  volume={15},
  number={2},
  pages={2121--2124},
  year={2020},
  publisher={IEEE}
}

@article{wymeersch2020radio,
  title={Radio localization and mapping with reconfigurable intelligent surfaces: Challenges, opportunities, and research directions},
  author={Wymeersch, Henk and He, Jiguang and Denis, Benoit and Clemente, Antonio and Juntti, Markku},
  journal={IEEE Vehicular Technology Magazine},
  volume={15},
  number={4},
  pages={52--61},
  year={2020},
  publisher={IEEE}
}

@INPROCEEDINGS{10463937,
  author={Sahoo, Deepak Kumar and Chakraborty, Chanchal and Ruchi and Kundu, Debidas and Patnaik, Amalendu and Chakraborty, Ajay},
  booktitle={2023 IEEE Microwaves, Antennas, and Propagation Conference (MAPCON)}, 
  title={A 1-Bit Coding Reconfigurable Metasurface Reflector for Millimeter Wave Communications in E-Band}, 
  year={2023},
  volume={},
  number={},
  pages={1-4},
  doi={10.1109/MAPCON58678.2023.10463937}}

@article{cassidy2009ground,
  title={Ground penetrating radar data processing, modelling and analysis},
  author={Cassidy, Nigel J and Jol, HM},
  journal={Ground penetrating radar: theory and applications},
  pages={141--176},
  year={2009},
  publisher={Elsevier Amsterdam}
}

@article{ram2010simulation,
  title={Simulation and analysis of human micro-Dopplers in through-wall environments},
  author={Ram, Shobha Sundar and Christianson, Craig and Kim, Youngwook and Ling, Hao},
  journal={IEEE Transactions on Geoscience and remote sensing},
  volume={48},
  number={4},
  pages={2015--2023},
  year={2010},
  publisher={IEEE}
}

@article{li2021multiple,
  title={Multiple targets localization behind L-shaped corner via UWB radar},
  author={Li, Songlin and Guo, Shisheng and Chen, Jiahui and Yang, Xiaqing and Fan, Shihao and Jia, Chao and Cui, Guolong and Yang, Haining},
  journal={IEEE Transactions on Vehicular Technology},
  volume={70},
  number={4},
  pages={3087--3100},
  year={2021},
  publisher={IEEE}
}
\bibliographystyle{ieeetr}
\end{document}